\documentclass[12pt]{article}
\usepackage[dvips]{graphicx}
\usepackage{quant}
\title{Qubit-qunit decomposition of quantum channel}
\author{Constantin V. Usenko}
\date{}
\begin{document}
\maketitle
\begin{abstract}
    Problem of classification of parallel quantum channels for information transfer is studied by method of ladder operators. 
   	Detailed compared to http://arxiv.org/abs/quant-ph/0702076 presentation of method of ladder operators is given. Examples of paired channel and qubit-qunit pair are added. Relations between eigenvalues of density matrices of composite channel and its subchannels determining presence/absence of entanglement of arbitrary mixed state of channel are obtained.
\end{abstract}
\tableofcontents

\section{Introduction}

Classical theory of information is based on possibility of decomposition of arbitrary message to finite sequence of bits. For states of quantum channels of information transfer entanglement –- impossibility of representation of state through direct product of states of subchannels is specific, this prevents decomposition of messages. Instead of that states of composite quantum channel are usually superpositions of such direct products (pure states), or mixtures of the superpositions. 

Properties of each specific state as to decomposition to subchannels depend of numerous factors:
on the one hand, for each pure state one can choose method for decomposition of channel to subchannels such that given state is represented with direct product of states of subchannels; on the other hand the whole set of possible states of composite quantum channel at each decomposition to subchannels includes more entangled states than pure states; besides that in classical approximation entanglement as the property of state of composite channel has to vanish completely. 

Study of entanglement has started as early as at time of formation of quantum physics (EPR paradox) and it is continued till now as one of the most fundamental parts of quantum information theory. Measures and characteristics of entanglement developed, besides intuitive idea of concurrence \cite{symbqs} and partial transposition \cite{Horodecki96}, are lately re-formulated to physically meaningful properties of covariance matrix \cite{Kothe}. 
Method of ladder operators in application to composite quantum channel \cite{Gdansk, VU1e, VU2e} is effective for investigation of the properties of density matrices of channel and its subchannels representing entanglement of state of channel.

\section{Ladder operators and algebra of observables}

Finite-dimensional state spaces with practically arbitrary dimensionality in quantum-mechanical problems are studied good enough –- all of those are unitarily equivalent to spaces in which irreducible representations of group of isotropy of three-dimensional space characterized with given value of momentum \(j=\frac{N-1}{2}\) are realized. Eigenvectors of operator $\hat{J}_{3}$ of momentum projection to given axis form basis associated to ladder operators $\hat{J}_{a}$. Ladder operators are expansion to multi-dimensional case of Stokes operators used recently for analysis of entanglement \cite{Cirac08}. Basis vectors are generated as result of sequential effect of one of ladder operators $\hat{J}_{\pm}$ on annihilating vector \(\cat{\mp j}\) of the other $\hat{J}_{\mp}\cat{\mp j}=0$:  
\[\cat{m} \propto \hat{J}_{+}^{m + j}\cat{- j}\propto \hat{J}_{-}^{j-m}\cat{ j}.\]
It is important that ladder operators can be constructed for each orthonormal basis \(\left\{\cat{k}:\ k=1\ldots N\right\}\) according to following definition:
\begin{equation}\label{ladder_def}
\begin{array}{l}
	\hat{J}_+\stackrel{Def}{=} \sum^{N}_{k=1}{\sqrt{\left(N-k\right)k}\cat{k+1}\otimes\bra{k}};\\
 \hat{J}_-\stackrel{Def}{=} \sum^{N}_{k=1}{\sqrt{\left(N+1-k\right)\left(k-1\right)}\cat{k-1}\otimes\bra{k}}.
\end{array}
\end{equation}
Operator 
\begin{equation}\label{ladder_z}
 \hat{J}_3\stackrel{Def}{=} \big(\hat{J}_+\hat{J}_--\hat{J}_-\hat{J}_+\big)/2= \sum^{N}_{k=1}{\left(k-\frac{1+N}{2}\right)\cat{k}\otimes\bra{k}}
\end{equation}
has basis vectors as eigenvectors. Re-definition of basis vectors by means of eigenvalues of operator $\hat{J}_{3}$ 
\[\cat{k}\rightarrow\cat{m=k-\frac{1+N}{2}}: m=-N/2\ldots N/2\] 
completes analogy between arbitrary basis and basis of irreducible representation of group \(SU(2)\). 

Ladder operators are the generating operators of matrix algebra in state space~–-~an arbitrary observable has representation with normally ordered combination of ladder operators:
\begin{equation}\label{O_norm}
	\hat{O}=\sum_{m\geq n=1}^N{\left(O_{m,n}+O^*_{n,m}\right)\hat{J}_+^m\hat{J}_-^n}.
\end{equation}
Each observable that is diagonal observable in chosen basis has eigenvalues $O_k$ that are interpolated with some function $o\left(x\right)$ of eigenvalues $k-\frac{1+N}{2}$ of operator $\hat{J}_3$ and thus it is the same function of that operator:
\begin{equation}\label{O_diag}
	\hat{O}=o\left(\hat{J}_3\right);\ \mapsto 
	\hat{O}=\sum_{\forall k}{o\left(k-\frac{1+N}{2}\right)\cat{k}\otimes\bra{k}}.
\end{equation}

To observable \(\hat{O}\) with non-degenerate spectrum \(\left\{O_k:\ k=1\ldots N\right\}\) basis formed with system of eigenvectors corresponds. To that basis such a system of ladder operators $\hat{J}^{\left\{O\right\}}_{a}$ corresponds that operator $\hat{J}^{\left\{O\right\}}_{3}$ commutates with that observable and representation
\[\hat{O} = O\left(\hat{J}^{\left\{O\right\}}_{3}\right);\ O\left(k-\frac{N+1}{2}\right)=O_k\]
takes place, for instance in form of interpolation Lagrange polynomial:
\begin{equation}\label{interpol}
	O\left(\hat{J}_3\right)=\sum_{k=1}^N{O_k\prod_{k\neq k'=1}^N{\frac{\hat{J}_3-k'-\frac{N+1}{2}}{k-k'}}}.
\end{equation}

\subsection{Preparation of states} In each specific event of information transfer quantum channel is prepared in one of pure states. Let us suppose that state system is orthogonal, general state of quantum channel is a weighted sum of states from that basis and each pure substate \(\roa{k}\) has weight set in process of preparation \(p_k\).   So, system of eigenstates of density matrix as well as eigenvalues of the matrix characterize the process of preparation of state. Sequence of states can be intentional if the channel is used for information transfer, it can be random in the case of control of channel states being absent or partially random if quality of control is not enough for total control of channel. In the last case one can expect that several states corresponding to one value of information being transferred are prepared with same probability.

General state is represented by diagonal density matrix with \(N\) set parameters \(p_k\) to which normalization condition is applied:
\begin{equation}
\hat{\rho} = \sum_{k=1}^N{p_k\roa{k}};\ \sum_{k=1}^N{p_k}=1.
\end{equation}
This matrix has representation with polynomial \(\rho\left(z\right)\) of degree \(N-1\) from ladder operator $\hat{J}_{3}$, that satisfies system of interpolation equations:
\begin{equation}
\hat{\rho} =  \rho\left(\hat{J}_{3}\right);\  \rho\left(k-\frac{N+1}{2}\right)=p_k.
\end{equation}

In arbitrary basis density matrix of quantum channel is not diagonal, it differs additionally by values \(N\left(N-1\right)\) of free parameters of transformation matrix from group \(SU(N)\), therefore in general density matrix in arbitrary basis of state space has  \(N^2-1\) free parameters: eigenvalues of density matrix and characteristics of basis of its eigenvectors. 

Information completeness of state can be calculated by formula for von Neumann entropy \[S_N=-\tr{\hat{\rho}\log_2 \hat{\rho}}.\]
Since in basis of eigenvectors density matrix is diagonal value of entropy is calculated explicitly:
\[S_N=-\sum_{k=1}^N{p_k\log_2 p_k}.
\]

\subsection{Measurement of states} Process of measurement of state of quantum channel of information transfer in each specific event is carried out in two stages. At first stage interaction of channel with analyzer and reduction of density matrix of system to mix of states determined by analyzer take place.

Mathematic representation of analyzer is orthogonal resolution of identity: complete system of orthogonal projectors \(\left\{\roa{k}:\ k=1\ldots N\right\}\)  projecting state after interaction with analyzer, or ladder operator \(\hat{J}_3\) related to those projectors.  The ladder operator can be respectively considered as operator of observable.

After choice of analyzer as result of measurement there can be either distribution of probabilities:
 \[p_k=\tr{\hat{\rho}\hat{J}_3},\ k=1\ldots N-1, \]
 that give $N-1$ real values: diagonal components of density matrix of measured state, or $N-1$ moments of observable: 
 \[\aver{\hat{J}^k_3}=\tr{\hat{\rho}\hat{J}^k_3},\ k=1\ldots N-1,\] 
 values of which make it possible to determine same diagonal components of density matrix. Since $N$-th moment of observable depends on previous $N-1$ moments such result corresponds to complete measurement for each observable compatible with analyzer, i.e. such one that its matrix has eigenvectors same to those of analyzer.

Arbitrary density matrix has $N^2-1$ independent real parameters, therefore one series of measurements can not determine state of system under consideration; an exception is observable, which basis of eigenvectors is identical to system of states in which quantum state of information transfer is prepared.  

In general case $N+1$ series of measurements are needed that use essentially different bases, i.e. such ones that do not transform to each other with simple permutation of vectors. For such bases operators  $\hat{J}_3^{\left\{m\right\}}$ have not to commutate one with another. Sequence of mutually non-commutative operators one can easily construct having defined sequence of angles $\phi^{\left\{m\right\}}=m\pi/(N+1)$. Sequence of $N+1$ operators: 
\begin{equation}\label{tomo}
	\hat{J}_3^{\left\{m\right\}}
=\cos{\phi^{\left\{m\right\}}}\hat{J}_3
+\sin{\phi^{\left\{m\right\}}}\left(\hat{J}_++\hat{J}_-\right)/2:\ m=0\ldots N
\end{equation}
that do not commutate one with another \begin{equation}
	\Big[\hat{J}_3^{\left\{m\right\}}\hat{J}_3^{\left\{n\right\}}\Big]=
	\sin{\phi^{\left\{m-n\right\}}}\left(\hat{J}_+-\hat{J}_-\right)/2
\end{equation}
has as eigenvectors sequence of needed bases, and results of measurements of each of observables $\hat{J}_3^{\left\{m\right\}}$ additionally define $N-1$ real numbers, independent characteristics of density matrix of measured state. General number of measurables coincides with number of real parameters of density matrix and in such way one obtains complete information of statistical properties of state of quantum channel.

The scheme described sets realization of method of quantum tomography in application to arbitrary quantum system with finite-dimensional state space. Its advantage compared to standard schemes \cite{Lidar08} is in coordination with properties of source and/or detector since ladder operators are coordinated with system of basis vectors generated with properties of specific quantum channel of information transfer.

 \section{Algebra of observables of paired channel}
 Hereinafter \(N=N_A\cdot N_B\) denotes dimensionality of state space \(\mathcal{H}=\mathcal{H}_A\otimes \mathcal{H}_B\) of composite (paired) channel of information transfer, \(N_A\) – of state space  \(\mathcal{H}_A\) of smaller subchannel, \(N_B\) – of state space \(\mathcal{H}_B\) of the larger one.  
 
 Observed values of paired channel are represented by matrices \(N\times N=N_A\cdot N_B\times N_A\cdot N_B\) in common state space, of first subchannel – by matrices  \(N_A\times N_A\), of the second one -- by \(N_B\times N_B\). Arbitrary observable of paired channel is some algebraic combination of subchannels observables.
  To arbitrary basis of state space of paired channel \(\left\{\catt{k}:\ k=1\ldots N\right\}\) ladder operators \(\hat{J}_a\) \eref{ladder_def} are associated.

  In induced basis \(\catt{m,n}=\cat{m}\otimes\catr{n}\) observables of composite channel are represented by matrices with double indices: 
\begin{equation}
	O_{mn,m'n'}=\matt{m,n}{\hat{O}}{m',n'};\ \hat{O}=\sum_{\forall n,m,n',m'}{O_{mn,m'n'}\pro{m,n}{m',n'}}.
\end{equation}
One can separate among observables of channel the observables hat depend on state of subchannel \(A\) or \(B\) only:
 \[	A_{mn,m'n'}=A_{m,m'}\delta_{n,n'};\ B_{mn,m'n'}=\delta_{m,m'}B_{n,n'}.\]
Invariant form of observables of subchannels:
\begin{equation}
\hat{A}=\sum_{\forall m,m'}{A_{m,m'}\ro{m}{m'}}\otimes \hat{I};\ 
\hat{B}= \hat{I}\otimes\sum_{\forall n,n'}{B_{n,n'}\ror{n}{n'}}.
\end{equation}

  \subsection{Ladder operators of subchannels} Let us denote, with similarity to rule of angular moment addition in mind, \(s=\frac{N_A-1}{2}\) and \(l=\frac{N_B-1}{2}\) the ranks of state spaces of subchannels, and put into correspondence to bases of subchannels \(\left\{\cat{m}\right\}\)  and \(\left\{\catr{n}\right\}\) sets of ladder operators.
\begin{itemize}
	\item Subchannel A:
\begin{equation}\label{ladder_defS}
\begin{array}{l}
	\hat{S}_+\stackrel{Def}{=} \sum^{N_A}_{m=1}{\sqrt{\left(N_A-m\right)m}\cat{m+1}\otimes\bra{m}};\\
 \hat{S}_-\stackrel{Def}{=} \sum^{N_A}_{m=1}{\sqrt{\left(N_A+1-m\right)\left(m-1\right)}\cat{m-1}\otimes\bra{m}}; \\
 \hat{S}_3\stackrel{Def}{=} \big(\hat{S}_+\hat{S}_--\hat{S}_-\hat{S}_+\big)/2= \sum^{N_A}_{m=1}{\left(m-\frac{1+N_A}{2}\right)\cat{m}\otimes\bra{m}};
\end{array}
\end{equation}
	\item Subchannel B:
\begin{equation}\label{ladder_defL}
\begin{array}{l}
	\hat{L}_+\stackrel{Def}{=} \sum^{N_B}_{n=1}{\sqrt{\left(N_B-n\right)n}\ecat{n+1}\otimes\ebra{n}};\\
 \hat{L}_-\stackrel{Def}{=} \sum^{N_B}_{n=1}{\sqrt{\left(N_B+1-n\right)\left(n-1\right)}\ecat{n-1}\otimes\ebra{n}}; \\
 \hat{L}_3\stackrel{Def}{=} \big(\hat{L}_+\hat{L}_--\hat{L}_-\hat{L}_+\big)/2= \sum^{N_B}_{n=1}{\left(n-\frac{1+N_B}{2}\right)\ecat{n}\otimes\ebra{n}}.
\end{array}
\end{equation}
\end{itemize}
Each of the sets separately is generating for algebra of observables of respective subchannel, along with that they realize irreducible representations of group \(SU(2)\), thus direct sum of those :
\begin{equation}\label{sub_ladder}
	\hat{J}^{\left\{ind\right\}}_a=\hat{S}_a\otimes \hat{I}+\hat{I}\otimes \hat{L}_a;
\end{equation}
realizes as well representation of the same group in state space of channel and generates induced ladder operators of paired channel. This representation is direct sum of irreducible representations of ranks \(\left\{l-s \ldots l+s\right\}\).

Eigenvectors \(\catt{j,m}\) of induced operator of ladder operator of composite channel\(\hat{J}^{\left\{ind\right\}}_3\)  are formed by rule of momentum adding:
\begin{equation}\label{clebsh}
	\catt{j,m} =\sum_{m_s=-s\ldots s}{C_{j,m;m_s}\cat{m_s}\otimes\catr{m-m_s}}.
\end{equation}
Here for Clebsch-Gordan coefficients denotation \(
	C_{j,m;m_s}=\bracatt{s,l,m_s,m-m_s}{j,m} \)
is used. 
 Induced ladder operators  \eref{sub_ladder} differ from ladder operators of channel  \eref{ladder_def} at least because of the fact that operator \(\hat{J}^{\left\{ind\right\}}_3=\hat{S}_3\otimes \hat{I}+\hat{I}\otimes \hat{L}_3\) is degenerate: part of its eigenvalues are same.

  Thus paired channel differs from one-part one by existence of two sets of ladder operators --  irreducible set \(\hat{J}_a\) of operators \eref{ladder_def} that are generating for algebra of observables of the channel in general and induced set \(\hat{J}^{\left\{ind\right\}}_a\)  formed by direct sums of ladder operators of subchannels \eref{sub_ladder}.

\subsection{Diagonalization of density matrices} 
States of paired quantum channel of information transfer are characterized by three density matrces:
\begin{itemize}
	\item Density matrix of composite channel: \(\hat{\rho}=\sum_{j=l-s}^{l+s}{\sum_{m=-j}^j{p_{j,m}\proa{j,m}}}\);
	\item Density matrix of subchannel \(A\): \(\hat{\rho}^{\left\{A\right\}}=
	\sum_{m_s=-s}^s{p_{m_s}\roa{m_s}}\);
	\item Density matrix of subchannel \(B\): \(\hat{\rho}^{\left\{B\right\}}=
	\sum_{m_l=-l}^l{p_{m_l}\eroa{m_l}}\).
\end{itemize}
In non-degenerate case each of those matrices has single set of eigenvectors and from  \eref{clebsh} it follows \cite{Gdansk, VU1e, VU2e}, that in case of system of vectors \eref{clebsh} being eigen system for density matrix of composite channel it is eigen system for density matrices of both subchannels.

Eigenvalues \(p_{j,m}\) of density matrix of composite quantum channel of information transfer determine probability of formation of respective state \(\catt{j,m}\) by channel source and are properties of not the channel but the source. Similarly, system of eigenvectors of density matrix is characteristic of source of channel since eigenvectors of density matrix are vectors of pure states formed by source in each separate event of information transfer by channel. Properties of medium of channel and its decomposition to subchannels determine the system of eigenvectors of subchannels and representation \eref{clebsh} of basis vectors of the channel.

Eigenvalues of density matrices of subchannels are:
\begin{equation}\label{PS}
		p^{\left\{A\right\}}_n=\sum_{j=l-s}^{l+s}{\sum_{m=-j}^{ j}{p_{j,m}C_{j,m;n}^2}};
\end{equation}
\begin{equation}\label{PL}
		p^{\left\{B\right\}}_k=\sum_{j=l-s}^{l+s}{\sum_{m=-j}^{ j}{p_{j,m}C_{j,m;m-k}^2}}.
\end{equation}
Those determine probabilities of registration of respective states of subchannels. 

To simultaneous measurement of both subchannels operation of some $m$-th detector of subchannel $A$ and $n$-th detector of subchannel $B$ simultaneousely corresponds. Result of simultaneous measurement of both subchannels is in joint probability distribution that can be calculated by density matrix of composite channel as trace of product of density matrix of paired channel and projectors to eigenstates of subchannels:
\begin{equation}
	P_{m,n}=\tr{\hat{\rho}\hat{P}^{\left\{A\right\}}_m\hat{P}^{\left\{B\right\}}_n}
	=\sum_{j=l-s}^{l+s}{p_{j,m+n}C^2_{j,m+n;m}}.
\end{equation}

\subsection{Properties of entangled states} Density matrix of paired quantum channel of information transfer is characterized in basis of eigenvectors $N$ by probabilities $p_k$ sum of which is equal to one, thus in basis of eigenvectors is $N-1=N_AN_B-1$ characteristics of state. Along with that density matrixes of subchannels are characterized with $N_A-1+N_B-1=N_A+N_B-2$ probabilities of eigenstates of density matrices of subchannels only. Thus, in general case density matrix can not be determined by results of independent measurements of density matrices of each subchannel. Exceptions are degenerate states for which density matrix is function of induced operator  \(\hat{J}^{\left\{ind\right\}}_3\):
\begin{equation}
	\hat{\rho}^{deg}=\rho\left(\hat{J}^{\left\{ind\right\}}_3\right).
\end{equation}
 
In fact the induced operator has $2\left(l+s\right)+1=N_A+N_B-1$ different eigenvalues only, and with account of condition for sum of eigenvalues this number completely coincides with number of eigenvalues of density of subchannels.
\begin{equation}\label{dia}
	\hat{\rho}_{sys}=\sum_{j=l-s}^{l+s}{\sum_{m=-j}^{j}{
	p_{j,m}\proa{j,m}
	}}.
\end{equation}

 \section{Measurement of states of paired channel}
Complete measurement of state of composite quantum channel of information transfer is to determine \(N^2-1\) real values – components of density matrix. Complete measurements of states of each of subchannels give \(N_A^2-1\)  and \(N_B^2-1\) real parameters -- and 
\(\left(N_AN_B\right)^2-1-\left(N_A^2-1\right)-\left(N_B^2-1\right)=\left(N_A^2-1\right)\left(N_B^2-1\right)\) characteristics of state remain not determined. 

Additional characteristics of state are components of covariance matrix \(Q=\aver{\hat{O}_A\otimes \hat{O}_B}
-\aver{\hat{O}_A}\aver{\hat{O}_B}\), that are to be measured in number equal for determination of all not determined parameters. Method of ladder operators makes it possible to construct system of observables needed for complete determination of all covariance matrix similar to the one used for method of quantum tomography \eref{tomo}.  

Let us denote angles \(\phi=\frac{2\pi}{N_B}\) and \(\psi=\frac{2\pi}{N_A}\). Systems of noncommuting in pairs operators:
\begin{equation}
	\hat{L}_3^{\left\{m\right\}}
=\cos{\phi^{\left\{m\right\}}}\hat{L}_3
+\sin{\phi^{\left\{m\right\}}}\left(\hat{L}_++\hat{L}_-\right)/2:\ m=0\ldots N_B,
\end{equation}
and \begin{equation}
	\hat{S}_3^{\left\{n\right\}}
=\cos{\psi^{\left\{n\right\}}}\hat{S}_3
+\sin{\psi^{\left\{n\right\}}}\left(\hat{S}_++\hat{S}_-\right)/2:\ n=0\ldots N_A,
\end{equation}
determine complete sets of observables of the method of quantum tomography for each of subchannels. Therefore the system   \(\left(N_A^2-1\right)\left(N_B^2-1\right)\) of components of covariance matrix 
\begin{equation}\label{tomo-pair}
{Q}_{m,n}=\aver{\hat{L}_3^{\left\{m\right\}}\otimes \hat{S}_3^{\left\{n\right\}}}
-\aver{\hat{L}_3^{\left\{m\right\}}}\aver{\hat{S}_3^{\left\{n\right\}}}.
\end{equation}

completes the observables of each of subchannels and makes it possible to determine completely state of composite quantum channel. By analogy with method of quantum tomography in application to separate quantum channel of information transfer, method for determination of state of paired quantum channel through measurement of coefficients of covariance matrix  \eref{tomo-pair} should also be named paired quantum tomography .

\subsection{Measurement of pure states} Pure states are the states with density matrix that has only one non-zero eigenvalue $\rho_{j,m}=1$
\begin{equation}\label{diag}
	\hat{\rho}_{j,m}=\proa{j,m}=\sum_{k,n=-s\ldots s}{C_{j,m;k}C_{j,m;n}
	\cat{m-k}\bra{m-n}\otimes\ecat{k}\ebra{n}
	}
\end{equation}
Result of measurement of state of one of subchannels is reduction of density matrix of state of composite channel in general and density matrix of the other subchannel. If the state of the channel is not direct product of states of subchannels measurement destroys pure state with its transformation into mix:
\[
\hat{\rho}_{j,m}=\sum_{m_s=-s\ldots s}{C_{j,m;m_s}^2
\roa{m-m_s}\otimes\eroa{m_s}};\] 
Matrices of state densities of subchannels\[
\hat{\rho}^{\left\{B\right\}}_{j,m}=\sum_{m_s=-s\ldots s}{C_{j,m;m_s}^2
\roa{m-m_s}};\] 
\[\hat{\rho}^{\left\{A\right\}}_{j,m}
=\sum_{m_s=-s\ldots s}{C_{j,m;m_s}^2
\eroa{m_s}}.
\]
are similar and can be represented by matrix\begin{equation}\label{sub_rho}
\hat{\rho}^{\left\{sub\right\}}_{j,m}=\sum_{n=-s\ldots s}{\rho_{j,m;n}
\roa{n}};\ \rho_{j,m;n}=C_{j,m;n}^2.
\end{equation}

Density matrix of state of subchannel \eref{sub_rho} has rank that is equal to dimension of the smaller of subchannels.

Correlation properties are determined by non-diagonal terms \eref{diag}, thus by terms of double sum with different indices $k\neq n$. 
Probability $P_{n}$ of registration of one of subchannels in state $n$ is equal:
\begin{equation}
	P_{n}^{\left\{A\right\}}=C_{j,m;m-n}^2;\ P_{n}^{\left\{B\right\}}=C_{j,m;n}^2.
\end{equation}
Probability $P_{j,m}\left(k,n\right)$ of simultaneous registration of one subchannels in state $k$ and the other one in state $n$ is equal:
\begin{equation}
	P_{j,m}\left(k,n\right)=C_{j,m;n}^2 \delta_{m-n,k},
\end{equation}
and conditional probability is equal: 
\begin{equation}
	P_{j,m}\left(k|n\right)=\delta_{m-n,k}.
\end{equation}

Thus states of subchannels are completely correlated if composite quantum channel of information transfer is in one of pure states\eref{diag}. 

There exist exactly two independent pure states for which statistical distribution of results of measurement is absent:

\begin{equation}
	\catt{j,\pm j}=\cat{s,\pm s}\otimes\catr{l,\pm l},
\end{equation}
 all the other pure states are entangled.

\subsection{Entropy of entanglement} 
Von Neumann entropy of composite quantum channel of information transfer is determined by function of density matrix of channel:
\begin{equation}\label{S_full}
	S_N=-\sum_{j,m}{p_{j,m}\log_2 p_{j,m}},
\end{equation}
and is characteristic of the method of preparation of states of channel. For pure state it is equal to zero, for equilibrium mix of all possible states it takes the largest value \(\log_2 N \). 

Degeneration of state is accompanied by increase of entropy. In fact, if two eigenvalues of density matrix almost coincide one can use substitution  \(p_1=p_m+d\), \(p_2=p_m-d\), then normalization condition

\[ p_1+p_2+ \sum_{k=3}{p_k} =2p_m+\sum_{k=3}{p_k}\] 
does not include input from difference of eigenvalues and therefore derivatives of entropy as functions of difference of eigenvalues are determined with two terms only: 
\[\frac{\partial S}{\partial d}=
\frac{\partial }{\partial d}\left(-\left(p_m+d\right)\log_2 \left(p_m+d\right)-\left(p_m-d\right)\log_2 \left(p_m-d\right)\right),\]
and are equal:
\begin{equation}
	\frac{\partial S}{\partial d}=\log_2 \frac{p_m-d}{p_m+d};\ \frac{\partial^2 S}{\partial d^2}=- \frac{2p_m}{p_m^2-d^2}.
\end{equation}
One can see that at degeneration \(d \mapsto 0\) first derivative tends to zero whereas second derivative remains negative.

Values of von Neumann entropy for subchannels are determined by distributions \eref{PS} and \eref{PL} of probabilities of registration of each of engenstates of subchannels: 
\begin{equation}\label{def_ss}
	S^{\left\{A\right\}}=-\sum_{m}{
	p^{\left\{A\right\}}_{m}\log_2 p^{\left\{A\right\}}_{m}};\ 
	S^{\left\{B\right\}}=-\sum_{m}{
	p^{\left\{B\right\}}_{m}\log_2 p^{\left\{B\right\}}_{m}}
	\end{equation}
Naturally, Shannon entropy for each of subchannels can differ due to inadequate choice of measuring devices.

Since even in the case of pure state \(\catt{j,M}\) of composite channel states of subchannels are usually mixed, conditional entropies \(S_{j,M}\) differ from zero and are determined by respective Clebsch-Gordan coefficients:
\begin{equation}\label{def_sp}
	S_{j,M}=	S^{\left\{A\right\}}_{j,M}=	S^{\left\{B\right\}}_{j,M}=-\sum_{m=-s\ldots s}{
	C_{j,M;m}^2\log_2 C_{j,M;m}^2}.
	\end{equation}
This expression determines entropy of entanglement \cite{VP,RHMH} of arbitrary pure state of composite quantum channel, and thus information on state of subchannel that can be obtained as result of measurement of state of the second subchannel. 

Value of entropy of entanglement is bounded from above with dimension of the smaller of the subchannels. Really, each of the sums \eref{def_sp} includes not more than \(2s+1\) terms, respectively to dimensionality of state space of the smaller subchannel, thus one has evaluation:
\begin{equation}
	S_{ent}\leq \log_2N_A=\log_2\left(2s+1\right).
\end{equation}

Entropy of entanglement \eref{def_sp} is generated with process of reduction due to measurement of pure state of paired channel. Input of that value to information capacity of paired channel depends on the method of measurement. 

In the case of measurement carried out only in one (the larger, for certainty) subchannel, given state of subchannel \(\cat{m_l}\) can be obtained if the state of the channel belongs to the set:
\[\left\{\catt{j,m_l+m_s}:\ m_s=-s\ldots s,\  j=\max\left(l-s,\left|m_l+m_s\right|\right)\ldots l+s\right\}.\]
 Number of those sets is somewhat smaller than \(\left(2s+1\right)^2=N^2_A\), thus the upper boundary of information on state of channel (Holevo limit) is equal: 
\[H=S_N-\sum_{j,m}{S_{j,M}} \geq S_N-2\log_2N_A.\]
For uniformly distributed channel the right-hand part is exactly equal to maximally possible value \(\log_2N_B\) of information obtained by the larger sub-channel.

In the case of measurement carried out independently in both channels (without account of correlation of results of measurements in subchannels)loss of entropy as result of measurement of pure state is \(2S_{j,M}\) and Holevo limit becomes smaller than the one admissible for the larger subchannel. 

Measurements with account of correlation between subchannels give incomplete determination of state as well since the same pair of eigenvalues  \(m_l\) and \(m_s\) of operators of observables of subchannels one can obtain for \(2s+1\) (or a bit less) different states \(\left\{\catt{j,m_l+m_s}:\ j=\max\left(l-s,\left|m_l+m_s\right|\right)\ldots l+s\right\}\). Respective Holevo limit can be estimated as \(S_N-\log_2N_A\).

 \section{Examples} General expressions of Clebsch-Gordan coefficients are not enough for explicit determination of all the properties of paired quantum channels of information transfer. Hereinafter examples demonstrating general properties of paired channel are studied in more detail.
 
 \subsection{Similar subchannels}   Let us suppose that paired quantum channel is divided into two parts with isomorphous state spaces \(\mathcal{H}_A=\mathcal{H}_B\). Respectively, state space of composite channel is space of recond rank tensors on states of each of the channels \(\mathcal{H}=\mathcal{H}_A\otimes\mathcal{H}_B=\mathcal{H}_A\otimes\mathcal{H}_A\).  Such a channel we call here a dual one.

  Eigenvectors of density matrix of dual channel are following linear combinations of vectors of induced basis:
\begin{equation}
	\catt{j,m}=\sum_{m_s=-l}^l{C_{j,m;m_s}\cat{m_s}\otimes\ecat{m-m_s}};\ m=-j\ldots j; j=0\ldots 2l.
\end{equation}

Density matrix of arbitrary mixed state: 
\begin{equation}
	\hat{\rho}=\sum_{j=0}^{2l}{\sum_{m=-j}^j{p_{j,m}\proa{j,m}}}.
\end{equation}
The state remains entangled till density matrix is not degenerate. To disentangled states density matrix correspond in which all the eigenvalues with same quantum number \(m\) are the same:
\begin{equation}
	p_{\left|m\right|,m}=p_{\left|m\right|+1,m}=\ldots =p_{2l,m};\ m=-2l\ldots 2l.
\end{equation}
 
 Now let us consider a more specific non-trivial example \(l=1\) (subchannels with smaller dimensionality are just a pair of qubits).
  
  The basis is formed by the following entangled state vectors:
  
Completely entangled state (\(j=0\)) includes all the basis vectors of state spaces of subchannels:  
\begin{equation}
\catt{0,0}=\frac{1}{\sqrt{3}}\cat{-1}\otimes\ecat{1}
-\frac{1}{\sqrt{3}}\cat{0}\otimes\ecat{0}+\frac{1}{\sqrt{3}}\cat{1}\otimes\ecat{-1},
\end{equation}
Entropy of entanglement of the state \(S_{0,0}=\log_23\).

Three partially entangled states (\(j=1\)) include only a pair of basis vectors of state spaces of subchannels each:  
\begin{equation}
\begin{array}{l}
\catt{1,-1}=\frac{1}{\sqrt{2}}\cat{-1}\otimes\ecat{0}
-\frac{1}{\sqrt{2}}\cat{0}\otimes\ecat{-1}\\	
\catt{1,0}=\frac{1}{\sqrt{2}}\cat{-1}\otimes\ecat{1}
-\frac{1}{\sqrt{2}}\cat{1}\otimes\ecat{-1}\\	
\catt{1,1}=\frac{1}{\sqrt{2}}\cat{1}\otimes\ecat{0}
-\frac{1}{\sqrt{2}}\cat{0}\otimes\ecat{1}\\	
\end{array}
\end{equation}
Entropy of entanglement of each of the states \(S_{1,m}=1\).

Three entangled states (\(j=2\)): two include only pair of basis vectors of state spaces of subchannels each whereas the third is completely entangled:  
\begin{equation}
\begin{array}{l}
\catt{2,-1}=\frac{1}{\sqrt{2}}\cat{-1}\otimes\ecat{0}+\frac{1}{\sqrt{2}}\cat{0}\otimes\ecat{-1}\\	
\catt{2,0}=\frac{1}{\sqrt{6}}\cat{-1}\otimes\ecat{1}
+\frac{2}{\sqrt{6}}\cat{0}\otimes\ecat{0}+\frac{1}{\sqrt{6}}\cat{1}\otimes\ecat{-1}\\	
\catt{2,1}=\frac{1}{\sqrt{2}}\cat{1}\otimes\ecat{0}+\frac{1}{\sqrt{2}}\cat{0}\otimes\ecat{1}\\	
\end{array}
\end{equation}
Entropy of entanglement of two states \(S_{2,\pm 1}=1\), of the third one -- \(S_{2,0}=\log_23-1/3\).
 
 Two not entangled states (\(j=2\)) include only one basis vector of state spaces of subchannels each:  \begin{equation}
\catt{2,-2}=\cat{-1}\otimes\ecat{-1};\  \catt{2,2}=\cat{1}\otimes\ecat{1}.	
\end{equation}
For uniformly distributed state \(p_k=1/9\)  and Holevo limit:
\[H=\log_29-\frac{5+3\log_23-1/3}{9},\]
 is equal to 3.88 bits instead of  4.755 bits for one-particle channel.
 
 \subsection{Bisection of composite channel}
Method of bisection is one of the most efficient methods of classical information theory since just with that method finite set of events is represented by subset of some power of Boolean algebra and points of event set  -- with sequence of zeroes and unities.

State space of quantum channel of information transfer of paired dimensionality at application of bisection method gets representation with direct product of two-dimensional state space of the smaller subchannel – qubit and \(\frac{N}{2} =2l+1\) -- dimensional state space of the larger subchannel. In the case of state space of composite channel having dimensionality \(2^d\) such separation of qubits can be continued \(d\) times till the state space of the second subchannel is left two-dimensional – decomposition of quantum channel of information transfer to multi-qubit takes place. In the case of state space of composite channel having other dimensionality it can always be embedded to space dimensionality of which is power of two and be considered as subspace of multi-qubit state space. 

In spite of state space separate state of quantum channel not always can be given by composition of qubits, that for it is needed for construction of channel to admit simultaneous measurement of each of qubits. So, quantum channel in which information carrier is one electron that can be in several potential wells is not a composite channel since concurrently only one measurable can be measured. Two electrons with arbitrary spin in two or more wells form a pair of qubits, and in four wells for decomposition into three qubits three electrons are needed.

Here properties of composite channel in which one of subchannels has two-dimensional state space, and the other one, arbitrary, are studied.
 
\subsubsection{Paragubit} The simplest example of such channel is a paraqubit that consists of two parts and has four-dimensional state of common states. That state space is divided into three-dimensional irreducible subspace (triplet one) with basis of eigenvectors of operator  \(J^{\left\{ind\right\}}_3\):
\[\catt{1,1}=\cat{1/2}\otimes\catr{1/2};\  \catt{1,-1}=\cat{-1/2}\otimes\catr{-1/2} \]
\[\catt{1,0}=\frac{1}{\sqrt{2}}\cat{1/2}\otimes\catr{-1/2}+\frac{1}{\sqrt{2}}\cat{-1/2}\otimes\catr{1/2}, \]
and one-dimensional (singlet) with basis:
\[\catt{0,0}=\frac{1}{\sqrt{2}}\cat{1/2}\otimes\catr{-1/2}-\frac{1}{\sqrt{2}}\cat{-1/2}\otimes\catr{1/2}. \]
Two states are non entangled, and two entangled states of paraqubit give for the particles same uniformly distributed mixed states with degenerate density matrices:
\begin{equation}
	\hat{\rho}^{\left\{A,B\right\}}=\frac{1}{2}\hat{1}.  
\end{equation}
In realization of paraquibit with pair of electrons asymmetry of states with respect to permutation decreases state space of composite channel to one-dimensional, always entangled. Realization with pair of photons has three-dimensional state space with one entangled and two not entangled states. Complete four-dimensional state space is possible in the case of use of different parts as one and the other qubits only. 

\subsubsection{Qubit-qutrit}The following example is pair of particles with six possible states. Result of separation of qubit is subchannel with three states – qutrit. Six-dimensional state space is separated into four-dimensional and two-dimensional irreducible subspaces.
Four-dimensional irreducible subspace of state space of composite channel has two not entangled states with basis vectors:
\[	\catt{3/2,3/2}=\cat{1/2}\otimes\ecat{1};\ 
	\catt{3/2,-3/2}=\cat{-1/2}\otimes\ecat{-1},\]
and two entangled ones\[
\begin{array}{l}
	\catt{3/2,1/2}=\sqrt{\frac{2}{3}}\cat{1/2}\otimes\ecat{0}+ \sqrt{\frac{1}{3}}\cat{-1/2}\otimes\ecat{1};\\
	\catt{3/2,-1/2}=\sqrt{\frac{1}{3}}\cat{1/2}\otimes\ecat{-1}+ \sqrt{\frac{2}{3}}\cat{-1/2}\otimes\ecat{0}.\\
	\end{array}
\]
Two-dimensional subspace has two entangled states:
\[
\begin{array}{l}
	\catt{1/2,1/2}=\sqrt{\frac{1}{3}}\cat{1/2}\otimes\ecat{0}- \sqrt{\frac{2}{3}}\cat{-1/2}\otimes\ecat{1};\\
	\catt{1/2,-1/2}=\sqrt{\frac{2}{3}}\cat{1/2}\otimes\ecat{-1}- \sqrt{\frac{1}{3}}\cat{-1/2}\otimes\ecat{0};\\
	\end{array}
\]

Entangled states have following one-particle density matrices:
\begin{equation}
	\begin{array}{lll}
	\catt{\frac{1}{2},\pm\frac{1}{2}}:& \frac{1}{3}\roa{0}+\frac{2}{3}\roa{\pm1};&
	\frac{1}{3}\eroa{\pm\frac{1}{2}}+\frac{2}{3}\eroa{\mp\frac{1}{2}}\\
	\catt{\frac{3}{2},\pm\frac{1}{2}}:& \frac{2}{3}\roa{0}+\frac{1}{3}\roa{\pm1};&
	\frac{2}{3}\eroa{\pm\frac{1}{2}}+\frac{1}{3}\eroa{\mp\frac{1}{2}}
	\end{array}
\end{equation}
In each of entangled pure states state density matrices of each of the parts of the qutrit have only two non-zero eigenvalues each, and there is no state equidistribution.

\subsubsection{Qubit in multi-state channel} Let us now consider bisection of composite channel with arbitrary given dimensionality \(N=2N_B\). Under consideration that the smaller subchannel is a qubit, i.e. has two-dimensional state space, quantum number of the larger subchannel is \(l=\frac{N_B-1}{2}\), and eigenvalues of the induced ladder operator \(J_3^{\left\{ind\right\}}\) are the following combinations of the vectors of induced basis:
\begin{equation}
\begin{array}{l}
	\catt{l+1/2,l+1/2}=\cat{1/2}\otimes\ecat{l}; \\  \catt{l+1/2,-l-1/2}=\cat{-1/2}\otimes\ecat{-l};\\
	m=-l+1/2\ldots l-1/2:\\
	\catt{l+1/2,m}=\sqrt{\frac{l+1/2+m}{2l+1}}\cat{1/2}\otimes\ecat{m-1/2}+ \sqrt{\frac{l+1/2-m}{2l+1}}\cat{-1/2}\otimes\ecat{m+1/2};\\
	\catt{l-1/2,m}=-\sqrt{\frac{l+1/2-m}{2l+1}}\cat{1/2}\otimes\ecat{m-1/2}+ \sqrt{\frac{l+1/2+m}{2l+1}}\cat{-1/2}\otimes\ecat{m+1/2}.
	\end{array}
\end{equation}

There are only two not entangled states, those correspond to combination of the end states of both subchannels.

Input of each of entangled states to density matrix of composite channel is given with the expressions:
\begin{equation}
	\begin{array}{l}
	\proa{l+1/2,m}=\\
	\frac{l+1/2+m}{2l+1}\proa{1/2,m-1/2}+
	\frac{l+1/2-m}{2l+1}\proa{-1/2,m+1/2}\\
	-
	\frac{\sqrt{\left(l+1/2+m\right)\left(l+1/2-m\right)}}{2l+1}\times\\
	\left(\pro{1/2,m-1/2}{-1/2,m+1/2}+\pro{-1/2,m+1/2}{1/2,m-1/2}\right);\\
		\proa{l-1/2,m}=\\
	\frac{l+1/2-m}{2l+1}\proa{1/2,m-1/2}+
	\frac{l+1/2+m}{2l+1}\proa{-1/2,m+1/2}\\
	+
	\frac{\sqrt{\left(l+1/2+m\right)\left(l+1/2-m\right)}}{2l+1}\times\\
	\left(\pro{1/2,m-1/2}{-1/2,m+1/2}+\pro{-1/2,m+1/2}{1/2,m-1/2}\right).\\
\end{array}
\end{equation}
Inputs from states with same quantum number \(m\) are same one to another. Those include with given coefficients diagonal in induced basis terms that generate similar terms in density matrices of subchannels:
\begin{equation}
	\begin{array}{l}
		\hat{\rho}^{\left\{A\right\}}_{l+1/2,m}=
	\frac{l+m+1/2}{2l+1}\roa{1/2}+	\frac{l+1/2-m}{2l+1}\roa{-1/2};\\
	\hat{\rho}^{\left\{A\right\}}_{l-1/2,m}=
	\frac{l+1/2-m}{2l+1}\roa{1/2}+	\frac{l+1/2+m}{2l+1}\roa{-1/2};
\end{array}
\end{equation}
\begin{equation}
	\begin{array}{l}
	\hat{\rho}^{\left\{B\right\}}_{l+1/2,m}=
	\frac{l+m+1/2}{2l+1}\eroa{m-1/2}+	\frac{l+1/2-m}{2l+1}\eroa{m+1/2};\\
	\hat{\rho}^{\left\{B\right\}}_{l-1/2,m}=
	\frac{l+1/2-m}{2l+1}\eroa{m-1/2}+	\frac{l+1/2+m}{2l+1}\eroa{m+1/2}.
\end{array}
\end{equation}

Inputs to density matrix of composite channel from cross terms
\[	\begin{array}{l}
	\pm
	\frac{\sqrt{\left(l+1/2+m\right)\left(l+1/2-m\right)}}{2l+1}\times\\
	\left(\pro{1/2,m-1/2}{-1/2,m+1/2}+\pro{-1/2,m+1/2}{1/2,m-1/2}\right)\\
\end{array}
\]
  differ with sign only. 
  
  Just the cross terms are responsible for entanglement of states, thus state of composite channel is disentangled only in the case of compensation of all the cross terms, this is provided if all the differences of eigenvalues \(p_{l+1/2,m}-p_{l-1/2,m}\) go to zero:
\begin{equation}
	q_m=p_{l+1/2,m}-p_{l-1/2,m}; m=-l+1/2\ldots l-1/2.
\end{equation}
Reduction to zero of all the differences is condition of disentanglement of arbitrary mixed state. 

Density matrices of subchannels are diagonal 
\[\hat{\rho}^{\left\{A\right\}}=\frac{1}{2}\hat{I}+\frac{q}{2}\hat{\sigma}_z;\ 
\hat{\rho}^{\left\{B\right\}}=\sum_{\mu=-l}^l{	p_\mu\eroa{\mu}}
\]
Density matrix of the smaller subchannel is characterized with one parameter \(q\) – deflection from equilibrium state:
\begin{equation}\label{rule_q}
	q=\sum_{m=-l+1/2}^{l-1/2}{\frac{2m}{2l+1}q_m}+\frac{1}{2}p_{l+1/2,l+1/2}-\frac{1}{2}p_{l+1/2,-l-1/2},
\end{equation}
therefore weighted average value of parameters of entanglement \(q_m\) for each pair of modes is equal 
\begin{equation}\label{rule_qs}
	\sum_{m=-l+1/2}^{l-1/2}{\frac{m}{2l+1}q_m}=q-\frac{1}{2}p_{l+1/2,l+1/2}+\frac{1}{2}p_{l+1/2,-l-1/2}.
\end{equation}
This sum is not equal to zero if at least some cross terms are not compensated and the state is entangled.
 
Eigenvalues of density matrix of the larger subchannel are determined by combinations of four adjacent eigenvalues of density matrix of the channel:
\[	p_\mu=	\frac{l+1+\mu}{2l+1}p_{l+1/2,\mu+1/2}+\frac{l+1+\mu}{2l+1}p_{l+1/2,\mu-1/2}+\]
		\[\frac{l+1+\mu}{2l+1}p_{l+1/2,\mu+1/2}+\frac{l-\mu}{2l+1}p_{l-1/2,\mu-1/2},
\]
and have following representation:
\begin{equation}\label{rule_sum}
		p_\mu=p_{l+1/2,\mu+1/2}+p_{l+1/2,\mu-1/2}-\frac{l-\mu}{2l+1}q_{\mu+1/2}-\frac{l+\mu}{2l+1}q_{\mu-1/2}.
\end{equation}
Disentanglement of all the pairs of states of composite channel with same values \(m\) is accompanied with transformation of density matrix of qubit to equilibrium one. At that density matrix of the larger channel remains almost arbitrary: its eigenvalues become sums of neighboring eigenvalues of density matrix of composite channel:
\begin{equation}\label{rule_sum2}
		p_\mu=p_{l+1/2,\mu+1/2}+p_{l+1/2,\mu-1/2};\ \mu=-l \ldots l,
\end{equation}
though one should not expect any additional degeneration independent of properties of density matrix of not entangled composite channel.

Sequence of relations \eref{rule_sum} can be used as indication of entanglement of mixed state of composite channel. In fact, each non-zero value in sequence:
\begin{equation}\label{rule_dif}
	dp_\mu=	p_\mu-p_{l+1/2,\mu+1/2}-p_{l+1/2,\mu-1/2};\ \mu=-l \ldots l,
\end{equation}
makes evidence of absence of disentanglement of one of \(\left[l-1/2\right]\) pairs of states.

\paragraph{Indications of entanglement of qubit in composite channel. }
As the result, study of bisection of composite quantum channel of information transfer carried out by method of ladder operators has the following indications of entanglement of qubit state with the state of the second cubchannel:
\begin{itemize}
	\item deflection from equilibrium of density matrix of qubit, according to formula \eref{rule_qs}, is indication of entanglement, though absence of such deflection is not indication of disentanglement of state;
	\item absence of \(\frac{N}{2}-1\) pairs of same eigenvalues of density matrix of composite quantum channel of information transfer is indication of entanglement of state;
	\item for each four eigenstates of density matrix of composite quantum channel that make input to some state \(\eroa{\mu}\) of the larger subchannel the expression \eref{rule_dif} makes it possible to check absence or presence of effect of entanglement of those states on the state of the channel.
\end{itemize}

\subsection{Entropy of composite channel with qubit} 
\begin{figure}[ht]
\centering
	\includegraphics{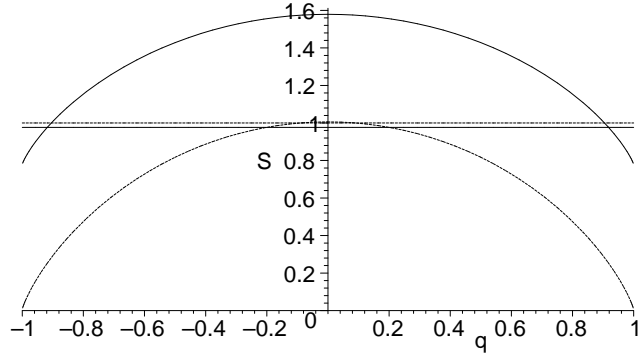}
	\caption{\label{S_bitQ}Dependence of entropy of qubit pair on difference of portions of triplet and singlet states. Firm lines show mixed state with noticeable admixture of not entangled states (\(r=0.9,d=0.6\)), dotted lines – mix of triplet and singlet states. Left edge of the graph corresponds to singlet state, the right edge – to triplet one.}
\end{figure}

Let us begin from the simpliest example.
Entropy of arbitrary mixed state of pair of qubits is determined by eigenvalues of density matrix: 
\begin{equation}\label{full_pq}
	\hat{\rho}=p_s	\hat{\rho}_s + p_{0}\proa{0,0}+ p_{1}\proa{1,1}+ p_{t}\proa{t}
\end{equation}
with portion $p_s$ of singlet state, portions $p_{0}$ and $p_{1}$ of states of independent parts and portion $p_{t}$ of entangled triplet state. Probabilities of states satisfy the conditions
\[
	p_s	=1- p_{0}- p_{1}- p_{t};\ p_{0}\geq0;\ p_{1}\geq0;\ p_{t}\geq0;\ p_s\geq0.
\]
In induced basis that very matrix looks like \begin{equation}\label{full_pqs}
\begin{array}{l}
	\hat{\rho}=	p_{0}\roa{0}\otimes\eroa{0}+p_{1}\roa{1}\otimes\eroa{1}\\
	 +	\frac{p_{t}+p_s}{2}\big(\roa{0}\otimes\eroa{1}+\roa{1}\otimes\eroa{0}\big)\\
+	\frac{p_{t}-p_s}{2}\big(
\cat{0}\bra{1}\otimes\ecat{1}\ebra{0}+
\cat{1}\bra{0}\otimes\ecat{0}\ebra{1}
\big)
\end{array},
\end{equation}
where one can easily see the part responsible for entanglement of states (the last line). This part vanishes in the case of coincidence of portion of singlet and entangled triplet states, $p_s=p_{t}$. Let us denote:
\[p_0=\frac{1-d}{2}\frac{1-r}{2};\ p_1=\frac{1-d}{2}\frac{1+r}{2};\ 
p_t=\frac{1+d}{2}\frac{1-q}{2};\ p_s=\frac{1+d}{2}\frac{1+q}{2}; \]
\[h\left(x\right)=-\frac{1+x}{2}\log_2\frac{1+x}{2}-\frac{1-x}{2}\log_2\frac{1-x}{2}.\]
In those denotations entropy of arbitrary state of pair of qubits is: 
\[S=h\left(d\right)+\frac{1-d}{2}h\left(r\right)+\frac{1+d}{2}h\left(q\right),\]
while entropy of one of qubits is given by expression:
\[S=h\left(\frac{1-d}{2}r\right).\]
If the portion \(\frac{1+d}{2}\) of entangled states is large enough, entropy of each of qubits exceeds entropy of composite state, as one can see from fig. \ref{S_bitQ}. The upper solid curve and the upper dotted curve represent dependence on parameter of mixing of entropy of state for pair of qubits. Horizontal solid line and dotted line represent entropies of each of the qubits of the pair. Till there is no admixture of not entangled states (\(d=1\)), entropy of each qubit separately is larger than entropy of pair. For such states process of measurement of one of qubits creates information on state of the other one. In the case of state of channel having noticeable admixture of one or both not entangled states of the pair entropy of one qubit can exceed entropy of pair only in the case of noticeable shift to one of entangled states.

Vanishing of entanglement of states in the example given above takes place due to degeneration of density matrix: coincidence of eigenvalues $p_s$ and $p_{t}$ corresponding to entangled states. Along with that coincidence of other pairs of eigenvalues does not bring to existence disentanglement of states.  

\paragraph{Qubit with qutrit.}
Arbitrary state of composition of qubit with qutrit is weighted mix of pure states and can be represented as
 \begin{equation}
	\begin{array}{r}
	\hat{\rho}_{p}= \sum_{m=\pm1/2}{p_{1/2,m}\proa{1/2,m}}\\
	+\sum_{m=\pm1/2,\pm3/2}{p_{3/2,m}\proa{3/2,m}}
\end{array}
\end{equation}
In induced basis
\begin{equation}
	\begin{array}{lr}
	\hat{\rho}_{p}&
	 =p_{3/2,+3/2}\roa{1}\otimes\eroa{1/2}+p_{3/2,-3/2}\roa{-1}\otimes\eroa{-1/2}\\
+&\left(\frac{1}{3}p_{3/2,1/2}+\frac{2}{3}p_{1/2,1/2}\right)\roa{1}\otimes\eroa{-1/2}\\
+&\left(\frac{1}{3}p_{3/2,-1/2}+\frac{2}{3}p_{1/2,-1/2}\right)\roa{-1}\otimes\eroa{1/2}\\
+&\left(\frac{2}{3}p_{3/2,1/2}+\frac{1}{3}p_{1/2,1/2}\right)\roa{0}\otimes\eroa{1/2}\\
+&\left(\frac{2}{3}p_{3/2,-1/2}+\frac{1}{3}p_{1/2,-1/2}\right)\roa{0}\otimes\eroa{-1/2}\\
+&
\left[\frac{\sqrt{2}}{3}
\left(p_{3/2,1/2}-p_{1/2,1/2}\right)
\left(\cat{1}\bra{0}\otimes\ecat{-1/2}\ebra{1/2}
+\cat{0}\bra{1}\otimes\ecat{1/2}\ebra{-1/2}\right)\right.
\\
+&\left.\frac{\sqrt{2}}{3}
\left(p_{3/2,-1/2}-p_{1/2,-1/2}\right)
\left(\cat{-1}\bra{0}\otimes\ecat{1/2}\ebra{-1/2}
+\cat{0}\bra{-1}\otimes\ecat{-1/2}\ebra{1/2}\right)
\right]
\end{array}
\end{equation}
In square brackets in two last rows terms responsible for entanglement of states are separated. There is possible partial ($p_{3/2,\pm1/2}=p_{1/2,\pm1/2};\ p_{3/2,\mp1/2}\neq p_{1/2,\mp1/2}$) or total ($p_{3/2,1/2}=p_{1/2,1/2};\ p_{3/2,-1/2}= p_{1/2,-1/2}$) disentanglement of states.

Among partially degenerate states the ones in which input of two basis states only is left are notable, for instance mix of states $\proa{3/2,3/2}$ and $\proa{3/2,1/2}$ with probabilities $(1\pm d)/2$.
\begin{equation}
	\begin{array}{r}
	\hat{\rho}_{p}=
+\frac{1}{2}\left(1-d/3)\right)\roa{1}\otimes\eroa{-1/2}
+\frac{1}{2}\left(1+d/3)\right)\roa{0}\otimes\eroa{1/2}\\
+\frac{\sqrt{2}}{3}d
\Big[
\cat{1}\bra{0}\otimes\ecat{-1/2}\ebra{1/2}
+\cat{0}\bra{1}\otimes\ecat{1/2}\ebra{-1/2}
\Big]
\end{array}
\end{equation}

\begin{figure}[ht]
\centering
	\includegraphics[width=3in]{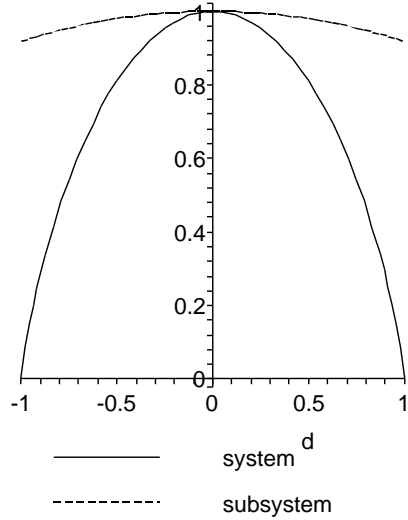}
	\caption{Dependence of entropy of paraqutrit and its qubit on non-degeneracy of state. }
	\label{fig}
\end{figure}
 Entropy of such state and its subchannels is
\[S_{sys}=\sth{\frac{1+d}{2}},
\]
\[S_{A}=S_{B}=\sth{\frac{1+d/3}{2}}.\]

To total degeneration of eigenvalues of density matrix value of non-degeneracy measure  $d=0$ corresponds. Graph in figure \ref{fig} represents dependence of entropy of paraqutrit and its qubit on measure of non-degeneracy of state. To completely degenerate state same value of entropy of paraqutrit and its qubit corresponds, while arbitrary deflection from degeneration is accompanied with entropy of qubit exceeding entropy of channel.

\paragraph{Qubit with qunit.} For non-degenerate states of channel entropy of composite channel is:
\begin{equation}\label{S_bi}
	S=-\sum_{m=-l-1/2}^{l+1/2}{p_{l+1/2,m}\log_2p_{l+1/2,m}}-\sum_{m=-l+1/2}^{l-1/2}{p_{l-1/2,m}\log_2p_{l-1/2,m}}.
\end{equation}
General expression for entropy of the smaller subchannel – qubit – can easily be obtained from the expression \eref{rule_q}:
\begin{equation}
	S_s=h\left(q\right),
\end{equation}
similarly the expression for entropy of the larger subchannel – from expression \eref{rule_sum}:
\begin{equation}
	S_s=-\sum_{\mu=-l}^l{p_\mu\log_2p_\mu}
\end{equation}

Effect of degeneration one can easily analyze in the case of bringing together two sums in \eref{S_bi}:
\begin{equation}\label{S_biu}
\begin{array}{l}
	S=-{p_{l+1/2,l+1/2}\log_2p_{l+1/2,l+1/2}}-{p_{l+1/2,-l-1/2}\log_2p_{l+1/2,-l-1/2}}\\
	-\sum_{m=-l+1/2}^{l-1/2}{\left(p_{l+1/2,m}\log_2p_{l+1/2,m}+p_{l-1/2,m}\log_2p_{l-1/2,m}\right)}.
	\end{array}
\end{equation}
Degeneration of pair of eigenvalues \(p_{l-1/2,m}\mapsto p_{l+1/2,m}\) can take place irrespectively from all the other eigenvalues of density matrix and is accompanied with increase of entropy to conditional maximum.

\section{Conclusions}
Method of ladder operators gives detailed information on properties of composite quantum channels.

Measurement of state of paired quantum channels of information transfer requires determination of not only density matrices of each of subchannels but covariance matrix as well. The method of paired quantum tomography proposed \eref{tomo-pair} makes it possible to determine completely state of paired quantum channel of information transfer with measurement of minimal needed number of incompatible observables.  

Number \(\left(N_A+2\right)\left(N_B+2\right)-1\) of needed incompatible observables is large enough, therefore indications of entanglement of states that would require measurement of substantially smaller number of observables are needed. In practical example of paired channel one of subchannels of which is qubit the following properties of density matrices of subchannels that are indications of entanglement of state exist: 
\begin{itemize}
	\item deflection from equilibrium of density matrix of qubit, according to formula \eref{rule_qs}, is indication of entanglement, though absence of such deflection is not an evidence of disentanglement of state;
	\item absence of \(\frac{N}{2}-1\) pairs of same eigenvalues of density matrix of composite quantum channel of information transfer is indication of entanglement of state;
	\item For each four eigenstates of density matrix of composite quantum channel that make input to some state \(\eroa{\mu}\) of the larger subchannel the expression \eref{rule_dif} for eigenvalues of composite state and the larger subchannel makes it possible to check absence or presence of entanglement of that very states.
\end{itemize}


\begin{thebibliography}{10}
\def\selectlanguageifdefined#1{
\expandafter\ifx\csname date#1\endcsname\relax
\else\language\csname l@#1\endcsname\fi}
\ifx\undefined\url\def\url#1{{\small #1}}\else\fi
\ifx\undefined\BibUrl\def\BibUrl#1{\url{#1}}\else\fi
\ifx\undefined\BibAnnote\long\def\BibAnnote#1{}\else\fi
\ifx\undefined\BibEmph\def\BibEmph#1{\emph{#1}}\else\fi

\bibitem{symbqs}
\selectlanguageifdefined{english}
\BibEmph{Wootters~W.} Entanglement of formation of an arbitrary state of two
  qubits~// \BibEmph{Phys. Rev. Lett.} "---
\newblock 1998. "---
\newblock Vol.~80. "---
\newblock Pp.~2245--2248. "---
\newblock bipartite qubit states.

\bibitem{Horodecki96}
\selectlanguageifdefined{english}
\BibEmph{Horodecki~M., Horodecki~P., Horodecki~R.} Separability of mixed
  states: necessary and sufficient conditions~// \BibEmph{Physics Letters A}.
  "---
\newblock 1996. "---
\newblock Vol. 223. "---
\newblock Pp.~1--8.

\bibitem{Kothe}
\selectlanguageifdefined{english}
\BibEmph{Kothe~C., Bjo"rk~G.} Entanglement quantification through local
  observable correlations~// \BibEmph{Phys. Rev. A.} "---
\newblock 2007. "---
\newblock Vol.~75, no.~1. "---
\newblock P.~012336. \BibUrl{ http://arxiv.org/quant-ph/0608041}.

\bibitem{Gdansk}
\selectlanguageifdefined{english}
\BibEmph{Usenko~C.~V.} Production of information and entropy in measurement of
  entangled states~// Proc. of the NATO ARW "Quantum Communication and
  Security", 10-13 September 2006,. "---
\newblock Gdansk, Poland: 2007. "---
\newblock Pp.~206--214.

\bibitem{VU1e}
\selectlanguageifdefined{english}
\BibEmph{Usenko~C.} Classification of two-particle quantum channels of
  information transfer: part 1~// \BibEmph{Bulletin of University of Kyiv,
  Series: Physics \& Mathematics}. "---
\newblock 2007. "---
\newblock no.~1. "---
\newblock Pp.~415--425.

\bibitem{VU2e}
\selectlanguageifdefined{english}
\BibEmph{Usenko~C.} Classification of two-particle quantum channels of
  information transfer: part 2~// \BibEmph{Bulletin of University of Kyiv,
  Series: Physics \& Mathematics}. "---
\newblock 2007. "---
\newblock no.~2. "---
\newblock Pp.~358--366. "---
\newblock quant-ph/0702076.

\bibitem{Cirac08}
\selectlanguageifdefined{english}
\BibEmph{Muschik~C.~A., Polzik~E.~S., Cirac~J.~I.} Detecting entanglement in
  two mode squeezed states by particle counting. "---
\newblock 2008. "---
\newblock P.~6. \BibUrl{ http://arxiv.org/quant-ph/0806.3448}.

\bibitem{Lidar08}
\selectlanguageifdefined{english}
\BibEmph{Mohseni~M., Rezakhani~A.~T., Lidar~D.~A.} Quantum-process tomography:
  Resource analysis of different strategies~// \BibEmph{Phys. Rev. A}. "---
\newblock 2008. "---
\newblock Vol.~77. "---
\newblock Pp.~032322--032337. \BibUrl{ quant-ph/0702131}.

\bibitem{VP}
\selectlanguageifdefined{english}
\BibEmph{Vedral~V., Plenio~M.} Entanglement measures and purification
  procedures~// \BibEmph{Phys.Rev. A}. "---
\newblock 1998. "---
\newblock Vol.~57. "---
\newblock Pp.~1619--1633. "---
\newblock quant-ph/9707035.

\bibitem{RHMH}
\selectlanguageifdefined{english}
\BibEmph{Horodecki~R., Horodecki~M., Horodecki~P.} Entanglement processing and
  statistical inference: The jaynes principle can produce fake entanglement~//
  \BibEmph{Phys. Rev. A}. "---
\newblock 1999. "---
\newblock Vol.~59. "---
\newblock Pp.~1799 -- 1803. "---
\newblock quant-ph/9709010[Issue 3 – March 1999].

\end{thebibliography}

\end{document}